\documentstyle[psfig,11pt]{article}
\def\nt{\hbox{$\nu_\tau$ }}

\newcommand{\AmS}{{\protect\the\textfont2
  A\kern-.1667em\lower.5ex\hbox{M}\kern-.125emS}}
\def\gsim{\;
\raise0.3ex\hbox{$>$\kern-0.75em\raise-1.1ex\hbox{$\sim$}}\;
}
\def\lsim{\;
\raise0.3ex\hbox{$<$\kern-0.75em\raise-1.1ex\hbox{$\sim$}}\;
}
\begin{document}
\thispagestyle{empty}
\begin{titlepage}
\begin{center}
\hfill FTUV/96-19\\
\hfill IFIC/96-21\\
\vskip 0.3cm
\LARGE
{\bf Nucleosynthesis constraints on heavy $\nu_{\tau}$ in the presence of
annihilations to majorons}
\end{center}
\normalsize
\vskip 1cm
\begin{center}
{\bf S.~Pastor
\footnote{Supported by Conselleria d'Educaci\'o i Ci\`encia of
Generalitat Valenciana. Work done in collaboration with J.C. Rom\~ao 
(IST, Lisbon), A.D. Dolgov and J.W.F. Valle (IFIC, Val\`encia).}\\
}
{\it{Instituto de F\'{\i}sica Corpuscular - C.S.I.C. \\
Departament de F\'{\i}sica Te\`orica, Universitat de Val\`encia \\
46100 Burjassot, Val\`encia, SPAIN}}\\[.15in]
Contribution to the {\it{IV International Workshop \\
on Theoretical and Phenomenological Aspects of \\
Underground Physics}}, Toledo, Spain, 17-21 September 1995.
\end{center}

\vskip 1cm
 
\begin{abstract}
We show that in the presence of sufficiently strong \nt 
annihilations to majorons, primordial nucleosynthesis constraints 
can not rule out any values of the \nt mass up the present laboratory
limit.
\end{abstract}
\end{titlepage}


The tau-neutrino is the only one which can have mass in the
MeV range. Among all studied limits on $m_{\nu_{\tau}}$, the
kinematical ones are the only model-independent bounds. The
last result from ALEPH \cite{Aleph95} is
\begin{equation}
m_{\nu_{\tau}} < 23~MeV
\end{equation}
In addition to this, there are stronger bounds from cosmological
considerations. From the contribution of stable $\nu_\tau$ to the
present relic density one gets $m_{\nu_\tau} < 92 \Omega h^2~eV$
\cite{boundCosm}, where $h=H_0/(100~kms^{-1}Mpc^{-1})$. This means 
that a massive $\nu_\tau$ with mass in the MeV range must be 
unstable with lifetimes smaller than the age of the Universe.

Moreover, if massive $\nu_\tau$'s are stable during nucleosynthesis 
($\nu_\tau$ lifetime longer than $\sim 100$ sec), one can constrain
their contribution to the total energy density, using the observed 
amount of primordial helium. This bound can be expressed through 
an effective number of massless neutrino species $N_\nu$. Using
$N_\nu < 3.4-3.6$, the $\nu_\tau$ mass range
\begin{equation}
0.5~MeV < m_{\nu_\tau} < 35~MeV
\label{cons1}
\end{equation}
has been excluded \cite{KTCS91,DI93}. This forbids all $\nu_\tau$ 
masses on the few MeV range. However, such masses are theoretically 
viable \cite{Teomnu}, and interesting for a possible solution of
the structure formation problem as described in \cite{SF}. 

It is possible to weaken the constraints on \nt masses of (\ref{cons1}), 
by adding new interactions of \nt beyond the standard ones. 
One may consider either that the $\nu_\tau$'s are 
unstable during nucleosynthesis \cite{unstable} or that they possess 
new channels of annihilation beyond the standard ones. The last case 
is what we have considered.

\begin{figure*}
\centerline{
\psfig{file=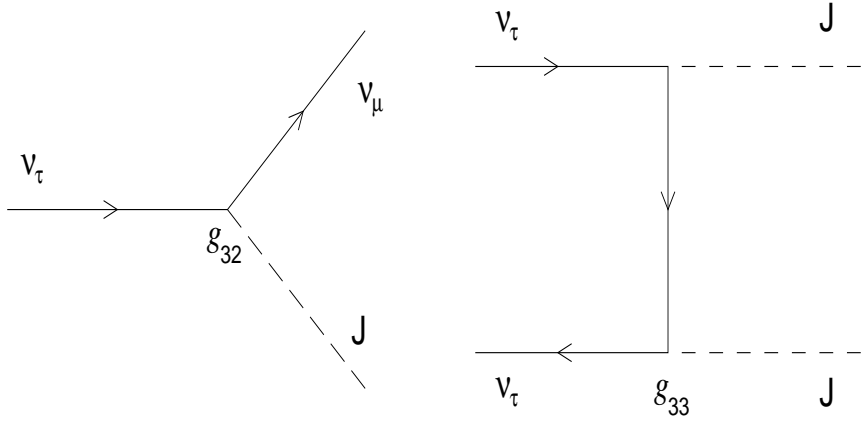,height=5.5cm,width=0.9\textwidth} }
\caption{Non-diagonal and diagonal couplings of Majorana tau
neutrinos and majorons.}
\label{couplings}
\end{figure*}

We studied the effect of the presence of Majorana $\nu_\tau$ 
annihilations to majorons (J). The interaction term of the
Lagrangian that describes the coupling between neutrinos
and majorons can be expressed \cite{Teomnu} as 
$\cal{L}_{\nu \nu J} \sim g J \nu^T \nu$, where $g$ 
is a $3\times 3$ matrix containing all possible couplings. 
In particular for $\nu_\tau$ the main processes are those 
described in figure (\ref{couplings}). We assume that
the coupling constants $g_{23}$ (non-diagonal) and $g_{33}$
(diagonal) are such that during nucleosynthesis time only 
annihilations are important. In previous works only the decay
processes were taken into account, though in general $g_{23}$ 
and $g_{33}$ may be simultaneously important.

We consider the interactions of massive $\nu_\tau$'s stable
during the epoch just before nucleosynthesis. The $\nu_\tau$'s 
interact with leptons via the standard weak interactions, $\nu_\tau
\bar{\nu}_\tau \leftrightarrow \nu_0 \bar{\nu}_0$, $e^+ e^-$, as 
in \cite{KTCS91,DI93}. We assume that, in addition, the $\nu_\tau$'s 
annihilate to majorons via the diagonal coupling
\begin{equation}
\cal{L} \sim gJ\nu_\tau^T \sigma_2 \nu_\tau
\end{equation}
where $g \equiv g_{33}$ and $\nu_\tau$ represents a two-component 
Majorana spinor. The corresponding elastic processes do not change
the particle densities, but as long as they are effective they 
maintain all species with the same temperature.

Even after the $\nu_\tau$'s have decoupled from standard weak 
interactions, they remain in contact with majorons. Therefore, 
as happens with photons and neutrinos when the $e^+ e^-$ pairs 
annihilate, after the decoupling there are two plasmas with 
different temperatures, one constituted of $\nu_\tau$'s and 
$J$'s and the other of the rest of the particles.

The evolution of the number densities of tau neutrinos ($n_\tau$)
and majorons ($n_J$) is given by  the solution of a set of 
Boltzmann differential equations
\begin{eqnarray}
\dot{n}_\tau + 3Hn_\tau = -\sum_{i=e,\nu_0} \langle\sigma_i v\rangle
(n_\tau^2-(n_\tau^{eq})^2) \nonumber \\
-\langle\sigma_J v\rangle(n_\tau^2-(n_\tau^{eq})^2
\frac{n_J^2}{(n_J^{eq})^2})
\label{Boltz1}
\end{eqnarray}
\begin{equation}
\dot{n}_J + 3Hn_J = \langle\sigma_J v\rangle(n_\tau^2-(n_\tau^{eq})^2
\frac{n_J^2}{(n_J^{eq})^2})
\label{Boltz2}
\end{equation}
where $\langle\sigma_\alpha v\rangle$ is the thermally averaged
cross section of the annihilation process of $\nu_\tau$'s
to particle-type $\alpha$ \cite{GonGel}, where
$\alpha=e,\nu_0,J$. In the above equations we 
assumed Boltzmann statistics and $n_i=n_i^{eq}$ for $i=e,\nu_0$.

Now let us briefly describe the calculations. First we normalized
the number densities to the number density of massless neutrinos,
$n_0 \simeq 0.18T^3$. We introduced $r_\alpha \equiv n_\alpha/n_0$,
where $\alpha=\nu_\tau,J$, and the
corresponding equilibrium functions $r_\alpha^{eq}$. On the 
other hand we performed the integrations using the dimension-less 
variable $x \equiv m_{\nu_\tau}/T$. The set of equations (\ref{Boltz1}) 
and (\ref{Boltz2}) is completed with the evolution of the $\nu_\tau$ 
temperature, which is obtained from one of Einstein's equations
\begin{equation}
\dot{\rho} = -3H(\rho + P)~.
\label{Einstein}
\end{equation}
Here $\rho$ and $P$ are the energy density and pressure of tau
neutrinos and majorons, respectively.

We have integrated numerically the coupled differential equations
in (\ref{Boltz1}-\ref{Einstein}), obtaining the solutions of $r_\tau$ 
for each pair of values ($m_{\nu_\tau},g$). The initial conditions are, 
for sufficiently high temperatures, $r_\tau = r_\tau^{eq}$, $r_J = r_J^{eq}$
and $T_\tau=T_J=T_{\nu_0}$.

The value of $r_\tau(m_{\nu_\tau},g)$ is used to estimate the
variation of total energy density $\rho_{tot}=\rho_{R} +
\rho_{\nu_\tau}$. In $\rho_{R}$ all relativistic species are taken
into account, including majorons and two massless neutrinos, whereas
$\rho_{\nu_\tau}$ is the energy density of massive $\nu_\tau$'s.

We can calculate now the effective number of massless neutrino
species ($N_\nu$) corresponding to each $r_\tau(m_{\nu_\tau},g)$. First
we run a program which calculates the evolution of the neutron fraction
($r_n$), as presented e.g. in \cite{rnevol}, varying the value of
$N_\nu$. Then we incorporate $\rho_{tot}$ to the same program and
perform the integration for each pair of values $(m_{\nu_\tau},g)$.
Comparing $r_n$ obtained in each case at $T_\gamma \simeq 0.065~MeV$
(the moment when practically all neutrons are wound up in $^4$He), we
can relate $(m_{\nu_\tau},g)$ and $N_\nu$.

\begin{figure}
\psfig{file=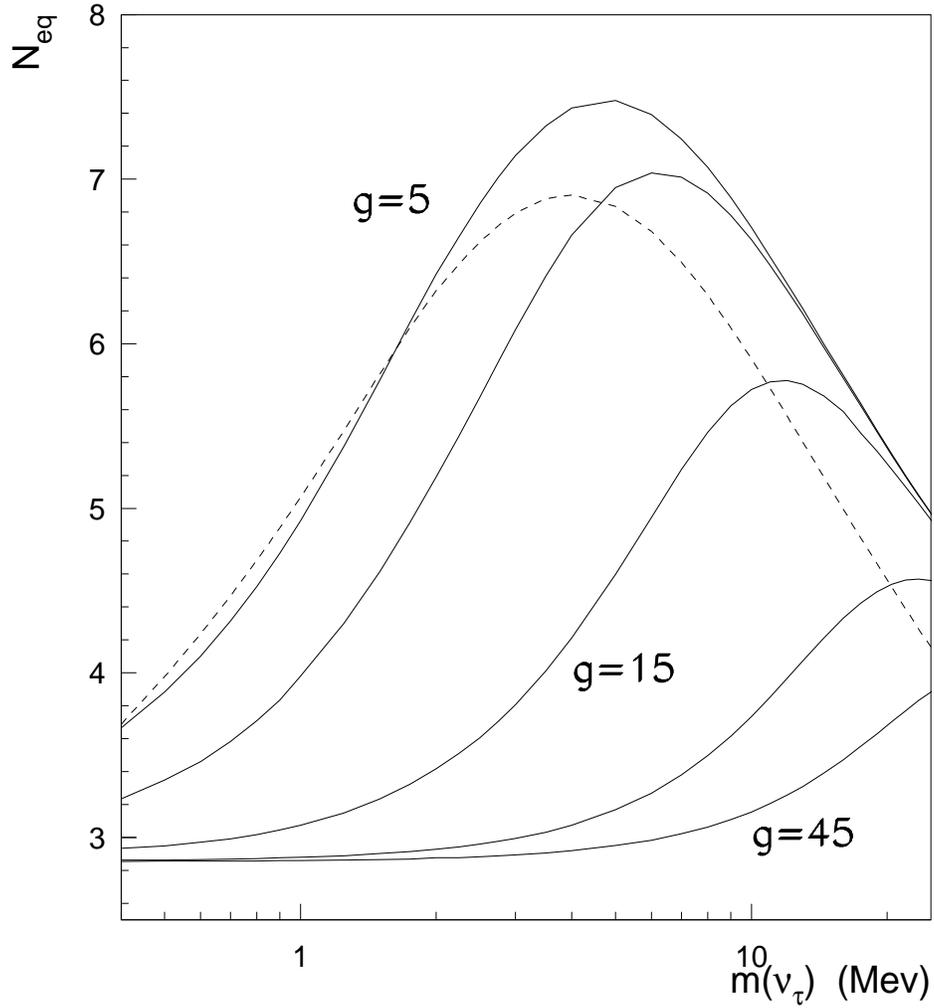,height=13.5cm,width=1.0\textwidth}
\caption{Effective number of massless neutrinos equivalent to
the contribution of massive neutrinos with different values of $g$ 
expressed in units of $10^{-5}$. For comparison, the dashed line
corresponds to the case when $g=0$ and no majorons are present.}
\label{neq}
\end{figure}

The results are shown in figure (\ref{neq}). One can see that for a
fixed $N_\nu^{max}$, a wide range of tau neutrino masses is
allowed for large enough coupling constants $g$. One sees that all
masses below 23 MeV are allowed by nucleosynthesis arguments, provided that the
coupling between $\nu_\tau$'s and $J$'s exceeds a value of a few
times $10^{-4}$. Such values are reasonable for many majoron models
\cite{Teomnu}. A more detailed study of this problem will be presented
elsewhere \cite{DPRV}.
%

Our conclusion is that the constraints on the mass of a Majorana
$\nu_\tau$ from primordial nucleosynthesis can be relaxed if
annihilations $\nu_\tau \bar{\nu}_\tau \leftrightarrow JJ$ 
are present.

\end{document}